\documentclass[conference,letterpaper]{IEEEtran}

\usepackage[utf8]{inputenc} 
\usepackage[T1]{fontenc}
\usepackage{url}
\usepackage{ifthen}
\usepackage{cite}
\usepackage{flushend}
\usepackage[cmex10]{amsmath} 

\usepackage{setspace}
\usepackage{balance}
\usepackage{xcolor}
\usepackage{amsthm}
\newtheorem{theorem}{Theorem}
\newtheorem{conjecture}{Conjecture}

\newtheorem{lemma}{Lemma}
\newtheorem{definition}{Definition}
\newtheorem{assumption}{Assumption}
\usepackage{cite}
\usepackage{amssymb,amsfonts}
\usepackage{algorithmic}
\usepackage{graphicx}
\usepackage{textcomp}
\usepackage{xcolor}
\usepackage{listings}
\usepackage{tikz}
\usepackage{main}
\usepackage{breqn}
\usetikzlibrary{positioning,chains,fit,shapes,calc}
\graphicspath{{./graphics/}}

\usepackage{setspace}

\begin{document}

\title{Coded Kalman Filtering over MIMO Gaussian Channels with Feedback} 

\author{%
  \IEEEauthorblockN{Barron Han, Victoria Kostina, Babak Hassibi}
  \IEEEauthorblockA{Department of Electrical Engineering\\
                    Caltech\\
                    Pasadena, CA, USA\\
                    Email: \{bshan, vkostina, hassibi\}@caltech.edu}
  \and
  \IEEEauthorblockN{Oron Sabag}
  \IEEEauthorblockA{School of Computer Science and Engineering\\
                    The Hebrew University of Jerusalem \\
                    Jerusalem, Israel\\
                    Email: oron.sabag@mail.huji.ac.il}
}

\maketitle

\begin{abstract}
We consider the problem of remotely stabilizing a dynamical system. A sensor (encoder) co-located with the system communicates with a controller (decoder), whose goal is to stabilize the system, over a noisy communication channel with feedback. To accomplish this, the controller must estimate the system state with finite mean squared error (MSE). The vector-valued dynamical system state follows a Gauss-Markov law with additive control. The channel is a multiple-input multiple-output (MIMO) additive white Gaussian noise (AWGN) channel with feedback. For such a source, a linear encoder, and a MIMO AWGN channel, the minimal MSE decoder is a Kalman filter. The parameters of the Kalman filter and the linear encoder can be jointly optimized, under a power constraint at the channel input. We term the resulting encoder-decoder pair a \emph{coded Kalman filter}. We establish sufficient and necessary conditions for the coded Kalman filter to achieve a finite MSE in the real-time estimation of the state. For sufficiency, we introduce a coding scheme where each unstable mode of the state is estimated using the channel outputs of a single sub-channel. We prove a coinciding necessity condition when either the source or channel is scalar and present a matrix-algebraic condition which implies the condition is necessary in general. Finally, we provide a new counter-example demonstrating that linear codes are generally sub-optimal for coding over MIMO channels.
\end{abstract}


 
\section{Introduction}
Controlling an unstable plant over a noisy communication channel is a hurdle for emerging technologies such as autonomous vehicles, Internet of Things devices, and remote surgery systems. This problem setting deviates from Shannon's communication problem in two ways that make it more challenging \cite{shannon}. First, the data to be transmitted correspond to physical measurements and arrive in a streaming fashion instead of being made available in its entirety before transmission, calling for causal encoders and decoders that operate in real time and maintain internal memory to improve the power-reliability tradeoff. Second, typical control systems are unstable, and their stabilization requires near-instantaneous and accurate estimates of the plant's state to produce effective control actions. Consequently, codes suitable for control systems must be low-delay yet highly reliable to perform control tasks over communication channels. We employ zero-delay joint source-channel codes to address the two objectives.

In a seminal paper, Sahai and Mitter \cite{sahaimitter} introduce the notion of \emph{anytime capacity}, the maximum transmission rate such that the decoding error decays exponentially with decoding delay. For scalar linear systems, they argue that a communication scheme capable of stabilizing a quantized version of the system can only exist if the logarithm of the system gain is less than the anytime capacity. However, computationally efficient coding schemes for this setting, especially joint-source channel codes, are generally unknown. Tree codes, as introduced by Schulman \cite{schulman}, achieve error probabilities that decay exponentially with the decoding delay. Although Sukhavasi and Hassibi showed a random construction of tree codes for a large class of discrete channels (even without feedback) \cite{sukhanytime}, to our knowledge, efficiently decodable tree codes have only been found for certain erasure channels \cite{sukhanytime}.

We consider a zero-delay joint source-channel coding problem where the goal is to design a computationally tractable encoder/decoder pair, separated by a noisy communication channel, so that the controller (decoder) can estimate the state of a linear dynamical system using outputs from the channel. Using these estimates of the source state, the controller provides control inputs to satisfy some control objective (i.e. stabilization).

The channel considered in this paper is a multiple-input multiple-output (MIMO) additive white Gaussian noise (AWGN) channel with feedback whose Shannon capacity is achieved by the water-filling solution \cite[Thm. 1]{telatarcapacity}. If the dynamical system to be controlled is directly connected to the channel without an input power constraint—--similar to the scenario in partially observed control—--the system can be stabilized given classical controllability and detectability conditions from control theory. A critical observation here is that because the noisy measurements of the system are linear combinations of the source states, any increase in the magnitude of the source states naturally leads to an increase in observation power, and consequently, a higher signal-to-noise ratio. In contrast, the MIMO AWGN communication channel model studied in this paper permits optimizing an encoder but imposes a fixed channel input power constraint, even when the source states become large.

A noiseless feedback channel connecting the decoder back to the encoder does not improve the Shannon capacity of a memoryless channel \cite{shannonfeedback}. Nevertheless, feedback can significantly simplify code design and improve the reliability-delay trade-offs for communication \cite{polyanskiy, burnashev}. Assuming a noiseless feedback channel is a reasonable idealization if the receiver has substantially more power than the transmitter, which is often the case in control systems, since the controller must provide essentially noiseless control inputs. Furthermore, noiseless feedback from the controller to the observer ensures that both the observer and the controller have access to the same history (``equi-memory"), which is a sufficient and necessary condition for separation of control and estimation to be optimal for linear systems and quadratic control cost \cite{sahaimitter}, \cite[Sec. IV]{kostinahassibirate}. Coding for bit-streaming sources over discrete channels, which is relevant to control systems where the state has been quantized for digital transmission, has been studied in \cite{nian2} for discrete memoryless channels with feedback and in \cite{sukhanytime} for the binary erasure channel without feedback. This indicates that codes with ``tree-like" properties \cite{schulman} may be simpler to implement for channels with feedback. This paper considers a setting where measurement and coding are analog operations applied in discrete time.

The source in this paper is a Gauss-Markov process. The causal rate-distortion function \cite{gorbunov} provides a lower bound to the channel capacity necessary to causally estimate the source over this channel subject to a given distortion \cite{kostinahassibirate, derpich, silva}. The causal rate-distortion function for the scalar Gauss-Markov process is known in closed form \cite{gorbunov} while that for the vector one is expressed as a semidefinite program \cite{Tanaka}. The causal rate-distortion function is lower bounded by the sum of logs of the Gauss-Markov process' unstable eigenvalues \cite{kostinahassibirate}. The latter quantity is identified as fundamental in a series of results from control literature known as data-rate theorems \cite{Nair2, elia, brockliber}. The lower bound to channel capacity provided by the causal rate-distortion function is known to be tight only if the source is matched to the channel at hand \cite{tocodeornot}. For example, a scalar Gauss-Markov source is matched to the scalar AWGN channel \cite{gorbunov, sahaimitter}, and a linear innovations encoder is optimal \cite{tatikonda, sahaimitter, Khina, elia}.

To achieve a finite MSE, the Shannon capacity should be greater than the sum of logs of unstable eigenvalues of the source \cite[Thm. 4.1]{yukselconverse}. For a vector source and parallel Gaussian channels with independent power constraints, \cite{ZAIDI201632} proposes a periodic linear encoder that, at each time, transmits a single source dimension over a single sub-channel, and shows sufficient conditions for achieving finite MSE. We obtain an equivalent sufficient condition using a time-invariant encoder and demonstrate the necessity of the condition in certain systems using techniques from linear estimation theory.

Classical results on controlling an unstable dynamical system under communication cost constraints include \cite{baillieul} and \cite{wong} that analyzed a scalar system, $\bfS_{t+1} = A \bfS_t + \bfU_t + \bfW_t$ with bounded initial state, $\bfS_0$, and bounded disturbances, $\{\bfW_t\}_{t\geq 0}$. The communication cost constraint is imposed as a noiseless bit-pipe that carries up to $r$ bits per time instance. It is determined in \cite{baillieul} and \cite{wong} that $r > \log A$ is a necessary and sufficient condition for the worst-case deviation of the system's state, in terms of the initial state and disturbances, to be finite at the infinite horizon. Tatikonda and Mitter generalize this result to the vector case, determining that the sum of logs of the unstable eigenvalues of $A$ must be less than the communication rate $r$ \cite{tatikondacontrol}. With an achievable scheme, Nair et al. showed that this condition is tight \cite{Nair}. If the disturbances $\bfW_t$ are stochastic and unbounded, i.e. Gaussian, Nair and Evans proved that the same condition is necessary and sufficient for the state to be mean-squared stabilizable: $\limsup_{t\to \infty} \E[\bfX_t^T \bfX_t] < \infty$ \cite{Nair2}. Other papers refine the problem setting to analyze the exact rate and control cost trade-offs \cite{kostinahassibirate, borkar, reducingminimum}.

Tight achievability conditions for stabilizing a system over noisy communication channels are rare. Stability over packet erasure channels was studied in \cite{sinopoliintermittent, sinopoli} using Kalman filter recursions with intermittent observations. Instead of encoding the source state as bits and transmitting using tree codes, an alternative is to perform measurement and coding as analog operations. To stabilize a scalar linear system over a binary symmetric channel, \cite{lalithakhinakostina} proposed a real-time posterior matching scheme and sufficient conditions to achieve $\eta$-moment stability. To stabilize the same system over an AWGN channel, \cite{Khina} considered a nonlinear encoder that maps a scalar linear system state to a space-filling curve transmitted over 2 channel uses and derived a sufficient condition for the joint source-channel code to achieve almost sure stability. The authors of \cite{continuous} derive necessary and sufficient conditions for stabilizing a continuous-time linear system over a continuous-time power-constrained infinite-bandwidth Gaussian channel.

This paper employs linear time-invariant codes to estimate a vector Gauss-Markov source over a MIMO AWGN channel with feedback. We show that the innovations' encoder that generates channel inputs as a linear function of the source estimation error (at the decoder) is optimal. The linearity of the encoder implies that the optimal decoder is a Kalman filter, and its MSE can be analyzed with linear estimation theory.


We identify sufficient and necessary conditions for a finite MSE to be achievable with linear codes for a given source, channel, and power constraint at the channel input. The analysis is carried out by showing an equivalence between achieving finite MSE and the existence of a stabilizing solution to a discrete algebraic Ricatti equation. The sufficient condition (achievability) requires each unstable source mode to be carried over a single MIMO sub-channel, thus the source modes are partitioned over the available sub-channels. This condition is shown to be necessary when either the source or channel is scalar. In particular, for a scalar source and a MIMO channel, it is shown that allocating the entire power to the least-noisy channel is optimal, while the typical water-filling solution that distributes the power among the subchannels is sub-optimal in general. The two solutions coincide for a sufficiently small power budget. To address the necessity of the condition in the scenario of a vector source over a MIMO channel, we put forth a matrix algebraic conjecture derived from the DARE feasibility condition, which, if true, implies the partitioning property is also necessary. We carried out numerical experiments that failed to disprove the conjecture.

Motivated by our achievable and converse results, we define a sufficient condition for source-channel matching in the stability sense, which guarantees that for certain source-channel pairs, linear codes can be utilized to achieve finite MSE with minimum power. We also demonstrate via an example that linear codes are not generally optimal.

The paper is organized as follows. Section \ref{setup} specifies the source and channel models and defines zero-delay feedback joint source-channel codes (JSCC) with an MSE performance criterion.
Section \ref{mainresult} presents our main contributions on the sufficient and necessary conditions for finite estimation error of a vector source over a MIMO Gaussian channel using linear codes and sketches our proof techniques. Our conditions coincide when either the source or channel is scalar. Section \ref{sec: control} casts our zero-delay JSCC problem as an equivalent remote control problem and translates our main results to the relevant linear quadratic regulator (LQR) problem. Specifically, separation of control and estimation implies finite LQR cost is achievable if and only if the source can be estimated with finite MSE. Finally, in Section \ref{sec:linnotopt}, we consider conditions for linear codes to stabilize a given source with minimal power compared to non-linear codes. We also demonstrate via a counterexample that linear codes are generally sub-optimal when the channel is not scalar. A preliminary version of this work was presented in \cite{barronisit}.

\section{Problem Setting and Preliminaries} \label{setup}
\subsection{Notation}
We denote by $\bfX_t, \ t \geq 0,$ a discrete-time random process and we denote by $\bfX^t$ the ordered collection of random vectors $\{\bfX_0, \bfX_1, \ldots, \bfX_t\}$. We write $\bfX \sim \mathcal N (\mu, \Sigma)$ to say that the random vector $\bfX$ has a Gaussian distribution with mean $\E[\bfX] = \mu$ and covariance matrix $\Cov[\bfX] = \Sigma$. Matrices and vectors are denoted with uppercase letters, while scalars are denoted with lowercase mathematical font. Sets are denoted in calligraphic font.

\subsection{System Model}
\begin{figure*}[!t]
  \centering
  \includegraphics[width=0.9\textwidth]{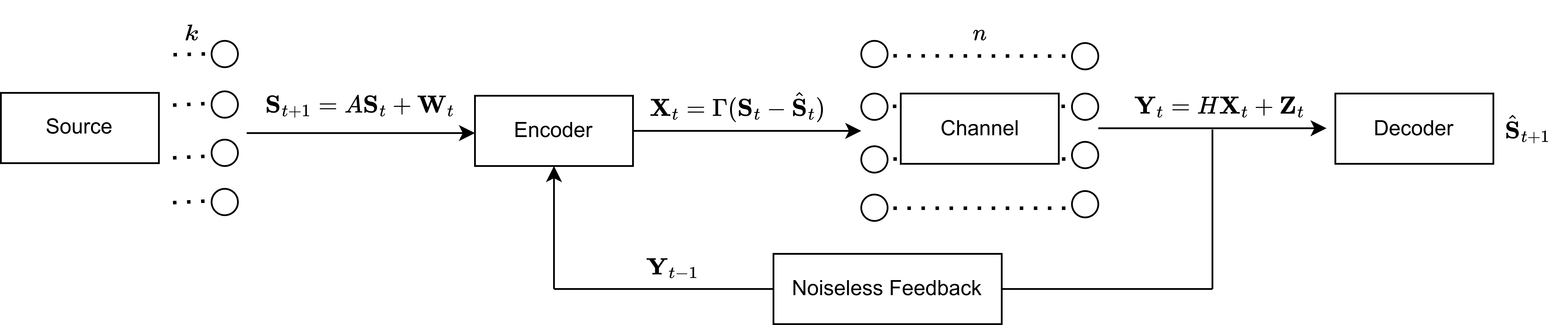}
  \caption{A MIMO AWGN channel, described in Definition \ref{channel}, with a noiseless feedback link is shown. The Gauss-Markov process, described in Definition \ref{source}, produces information at every time, which is encoded and passed through the channel. The decoder seeks to estimate the source at the next time given all channel outputs. We display the optimal encoding structure of Lemma \ref{optimalencoder}.}
  \label{fig: mimo}
\end{figure*}

Figure \ref{fig: mimo} depicts the setup. We define its main components: a MIMO AWGN channel, a Gauss-Markov streaming source, and a zero-delay code.

\begin{definition}[MIMO AWGN channel] \label{channel}
The AWGN channel, characterized by the deterministic gain matrix $H \in \mathbb{R}^{m \times n}$ and noise covariance $R \in \mathbb{R}^{m \times m}, \ R \succ 0$, accepts a vector input $\bfX_t \in \mathbb{R}^n$ and produces a vector output $\bfY_t \in \mathbb{R}^m$,
\begin{equation} \label{mimochannel}
    \bfY_t = H \bfX_t + \bfZ_t, \ t \ge 0.
\end{equation}
\end{definition}

We consider a channel with a diagonal channel matrix, $H = \mathrm{diag}\{h_1, \ldots, h_n\}$, $h_1 \geq h_2 \geq \ldots \geq h_n$, and identity noise covariance, $\bfZ_t \sim \mathcal{N}(0, I)$. This is without loss of generality, as any channel can be diagonalized to this form \cite[Thm. 1]{telatarcapacity}: for a channel given in Definition \ref{channel}, we whiten the noise $\mathbf {Z}$ by applying the transformation:
\begin{equation}
    R^{-\frac{1}{2}} \mathbf Y = R^{-\frac{1}{2}} H \mathbf{X} + R^{-\frac{1}{2}} \mathbf{Z}.
\end{equation}
We take a singular value decomposition $R^{-\frac{1}{2}} \tilde H = U \Sigma V^*$, where $U^*U = V^*V = I$ and $\Sigma$ is diagonal.
The equivalent diagonal channel is $\bar{\mathbf{Y}} = \Sigma \bar{\mathbf{X}} + \bar{\mathbf{Z}}$, with $\bar \bfX = V^* \bfX$, $\bar \bfY = U^* R^{-\frac{1}{2}} \bfY$, and $\bar \bfZ = U^* R^{-\frac{1}{2}} \bfZ \sim \mathcal{N}(0, I).$


The streaming source in Figure \ref{fig: mimo} is a Gauss-Markov source.
\begin{definition}[Gauss-Markov source]\label{source}
The Gauss-Markov source, characterized by the gain matrix $A \in \mathbb{R}^{k \times k}$ and noise covariance $Q \succ 0$, evolves according to:
\begin{equation} \label{stateupdate}
    \bfS_{t+1} = A \bfS_t + \bfW_t, 
\end{equation}
where $\bfW_t \stackrel{i.i.d.}{\sim} \mathcal{N}(0,Q)$ and the initial state is $\bfS_0 \sim \mathcal{N} (0, Q)$.
\end{definition}

Without loss of generality, we assume that the matrix $A$ can be written as
\begin{equation} \label{jordanA}
    A = \left[ \begin{array}{cc} A_s & 0 \\ 0 & A_u \end{array} \right],
\end{equation}
where $A_s$ is stable (eigenvalues on and inside unit circle), $A_u$ is strictly unstable (eigenvalues outside unit circle), and both $A_s, A_u$ are in Jordan form. This form can be accomplished by taking a Jordan decomposition of $A$ in Definition \ref{source}: $A = T^{-1} J T^{-1}$ where $J$ is block diagonal with the ordering given in \eqref{jordanA}. Then, we apply the state transformation $\tilde {\mathbf S}_{t} = T \mathbf{S}_t$, so that the evolution equation of $\tilde {\mathbf S}_{t}$ is given by $\tilde {\mathbf S}_{t+1} = J \tilde{\mathbf S}_t + \mathbf{W}_t.$

We make the following assumptions about the source:
\begin{assumption} \label{asm: controllable}
The pair $(A,Q)$ is controllable.
\end{assumption}
\begin{assumption} \label{asm: diag}
The source is strictly unstable, so $A_s = 0$, and $A_u$ in \eqref{jordanA} has distinct eigenvalues.
\end{assumption}

 Assumption \ref{asm: controllable} guarantees that the error covariance is positive definite \cite[Apx. C]{linearestimation}. Assumption \ref{asm: diag} yields a cleaner analysis since we can take $A$ to be diagonal when all eigenvalues are distinct as a consequence of the Jordan normal form of $A$. However, the results extend to any block\hbox{-}Jordan matrix $A$; see Appendix~\ref{app: jordanA} for details.
 We assume that the source is unstable without loss in generality since the stable part of the source can have a finite estimation error even if no communication is allowed. We limit the source to being strictly unstable so that there are no eigenvalues on the unit circle and a Lyapunov equation of the form $X = AXA^T + W$ has a unique solution \cite[Lem. D.1.1]{linearestimation}. This assumption is common in classical linear estimation theory \cite[Apx. C]{linearestimation}.
 
\begin{definition}[A zero-delay feedback JSCC]\label{code}
A $(p, D)$ feedback code for the source-channel pair in Definitions \ref{channel} and \ref{source} consists of the following:
\begin{enumerate}
  \item An encoder that at time $t$ has access to  $\bfS^t$ and $\bfY^{t-1}$ and generates
  \begin{equation} \label{ft}
      \bfX_t = f_t(\bfS^t, \bfY^{t-1}), \ t \geq 1,
  \end{equation}
  where
    $f_t \colon \mathcal{S}^t  \times \mathcal{Y}^{t-1} \mapsto \mathbb{R}^n$
    and $\mathcal S = \mathbb{R}^k, \mathcal{Y} = \mathbb{R}^m$.
    The channel inputs must satisfy an average power constraint,
    \begin{equation} \label{powerconstraint}
       \limsup_{T \to \infty} \frac{1}{T} \sum_{t=1}^{T} \E[\bfX_t^T \bfX_t] \leq p.
    \end{equation}
  \item A decoder, $g_t \colon \mathcal{Y}^{t} \mapsto \mathcal{S}$, that at time $t$ predicts the next source state,
  \begin{equation} \label{dec_def}
      \hat \bfS_{t} = g_t(\bfY^{t}).
  \end{equation}
\end{enumerate}
The long-term average mean-squared error must satisfy
\begin{equation} \label{distortion}
    \limsup_{T \to \infty} \frac{1}{T} \sum_{t=1}^{T}\mathbb{E}\left(\|\mathbf{S}_t - \hat{\mathbf{S}}_t\|^2\right) \leq D.
\end{equation}
\end{definition}

The code is zero-delay, in the sense that the estimate $\hat \bfS_{t+1}$ is produced by the decoder at time $t$ and the distortion criterion \eqref{distortion} incorporates the quality of every such estimate. The performance of the code is evaluated over an infinite time horizon $T \to \infty$, which helps us gain analytical insight into the achievable power-distortion tradeoff. The fundamental tradeoff of interest is that between available power $p$ and achievable distortion $D$, that is, for a given source and channel, one would like to characterize the set $\{p \geq 0, D \geq 0: \exists (p, D) \text{ feedback code}.\}$

Characterizing the tradeoff of achievable $(p, D)$ for the general class of encoders $\{f_t\}_{t=0}^\infty$ is challenging. Here, we limit our attention to encoders characterized by linear mappings $f_t$, which are more easily analyzed and implemented.
The zero-delay joint source-channel setting in Figure \ref{fig: mimo} and Definition \ref{code} is relevant to the control problem where the source in \eqref{source} is modified to include a control input $\bfU_t$,
\begin{equation} \label{lds}
    \bfS_{t+1} = A\bfS_t + B\bfU_t + \bfW_t,
\end{equation}
where $B$ is a constant matrix, and the decoder decides $\bfU_t$. Provided that $\bfU^t$ are available at the encoder at time $t$, the classical result \cite{introstoch} of separation holds, implying the system in \eqref{lds} is stabilizable, i.e. $\limsup_{t \to \infty} \E[\bfX_t^T\bfX_t] < \infty$ is achievable, if and only if the controller (decoder) can estimate the source with finite MSE, $D < \infty$ \eqref{distortion}. We extend this discussion in Section \ref{sec: control}.

\subsection{Optimal Linear Encoder}
A general linear encoder from Definition \ref{code} has the form 
\begin{equation} \label{encoding}
    \bfX_t = \Xi_t(\bfS^{t}) + \Psi_t (\bfY^{t-1}) + \bfM_t,
\end{equation}
where $\Xi_t, \Psi_t$ are linear mappings, and $\bfM_t$ is a Gaussian random variable independent of $(\bfS^t,\bfY^{t-1})$. This encoder involves all past states and channel outputs from feedback, and thus has input dimension that grows over time. The next result establishes a simplified optimal code structure involving only the recent state estimate error.

\begin{lemma}[Innovations encoder] \label{optimalencoder}
If a $(p, D)$ feedback code exists with an encoder of the form \eqref{encoding}, a $(p, D)$ feedback code exists with an encoder of the form
\begin{equation} \label{optencoding}
\bfX_t = \tilde \Gamma_t P_t^{-1} (\bfS_t-\hat{\bfS}_{t}) + \bfM_t,
\end{equation}
where $\bfM_t \sim \mathcal{N}(0, \Omega_t)$ is independent of $\bfS_t-\hat{\bfS}_{t}$, and $\hat{\bfS}_t$ is the optimal decoder's estimate, given recursively by
\begin{equation}\label{kalmandecode}
        \hat{\bfS}_{t+1} = \E[\bfS_{t+1} | \bfY^t] = A \hat{\bfS}_{t} + \mathrm K_{t+1} (\bfY_{t} - \tilde \Gamma_t H \hat \bfS_{t}),
\end{equation}
where $K_t = A\tilde \Gamma_t^T H^T \left( H \tilde \Gamma_t P_t^{-1} \tilde \Gamma_t^T H^T + I \right)^{-1}$, $\hat{\bfS}_0=0$, and the error covariance $P_t = \Cov(\bfS_t - \hat \bfS_t)$.
Encoder parameters $\tilde \Gamma_t, \Omega_t$ must satisfy the power constraint \eqref{powerconstraint}
\begin{equation}\label{powercons_sum}
    \limsup_{T\to\infty}\frac{1}{T} \sum_{t=1}^{T-1} \Tr(\tilde \Gamma_t P_t^{-1} \tilde \Gamma_t^T + \Omega_t) \leq p.
\end{equation}
\end{lemma}

\begin{proof}
    Let policy $U$ represent a general linear encoder of the form \eqref{encoding}. Consider an encoder $\bfX_t$ in \eqref{optencoding} and call it policy $V$. Note that $P_t$ is invertible since $Q,R \succ 0$ and $A$ is invertible by Assumption \ref{asm: diag} \cite[Lemma 9.5.1]{linearestimation}.  Let $\Theta^{U/V}_t \triangleq (\bfS_t - \hat \bfS_{t}, \bfX_t)^T$ be a Gaussian random vector induced by either policy $U$ or $V$. We show via induction that for any $\left\{ \Theta^U_t \right\}_{t<T}$, we can construct $\left\{ \Theta^V_t \right\}_{t<T}$ to have the same distribution. Specifically, set the parameters of the encoder in \eqref{optencoding} as 
    \begin{equation}
    \tilde \Gamma_t = \E_U\left[\bfX_t(\bfS_t - \hat \bfS_t)^T\right], \label{gt}
    \end{equation}
    and
    \begin{equation}
    \begin{aligned}
    \Omega_t
      &= \E_U[\bfX_t\bfX_t^T] \\
      &\quad - \E_U \bigl[\bfX_t(\bfS_t - \hat{\bfS}_t)^T\bigr] \,
              P_t^{-1} \,
              \E_U \bigl[(\bfS_t - \hat{\bfS}_t)\bfX_t^T\bigr].
    \end{aligned}\label{mt}
    \end{equation}
    This shows that the optimal prediction error covariance, $P_t$, using a general encoder can be achieved with a simplified policy $V$ at every time. The decoder estimate is given by the Kalman filter.
\end{proof}

Intuitively, the encoder can communicate what is currently unknown to the decoder with minimal power by transmitting the innovation $\bfS_t - \hat \bfS_{t}$. The decoder's prediction, $\hat \bfS_{t}$, is computed using channel feedback. Compared to \cite[Lem. 1]{barronallerton}, Lemma \ref{optimalencoder} above introduces an independent additive term $\bfM_t$ in \eqref{optencoding}. Although its covariance $\Omega_t$ can be chosen at will as long as \eqref{powercons_sum} is satisfied, we show in Lemma \ref{lemma: riccati}, below, that we can set $M_t = 0$, $\Omega_t = 0, \ \forall t$ if the goal is to minimize the MSE for a given power $p$. Encoders that utilize the innovations $\bfS_t - \hat \bfS_t$ are commonly used for low-delay coding of sources with memory over Gaussian channels. For example, \cite{oron, coverpombra, kim2} use a similar linear code to estimate the internal noise process of a colored MIMO Gaussian channel.

We make the following assumption on our encoder, which simplifies our analysis of the infinite-horizon estimation error.

\begin{assumption} \label{asm: time_inv_enc}
The encoder in Definition \ref{code} is time invariant which means $\tilde \Gamma_t = \tilde \Gamma, \Omega_t = \Omega, \forall t$ in \eqref{optencoding}.
\end{assumption}

The encoding structure in Lemma \ref{optimalencoder} reveals a state-space model, defined by \eqref{optencoding} and \eqref{mimochannel}, that admits a Kalman filter solution \eqref{kalmandecode}. Consequently, the Riccati recursions in Lemma \ref{lemma: riccati}, stated next, give the estimation error at the infinite horizon. This lemma generalizes Lemma 2 in \cite{barronallerton}.

\begin{lemma}[Riccati recursion and convergent behavior]\label{lemma: riccati}
For a fixed $\tilde \Gamma_t = \tilde \Gamma$, the prediction error covariance, $P_t = \Cov (\bfS_t - \hat \bfS_t)$,
of the Kalman filter designed for the estimation of the Gauss-Markov source \eqref{source} through an AWGN Channel \eqref{channel} whose input is formed as \eqref{optencoding} evolves according to a Riccati recursion that either diverges or converges to the stabilizing solution $P$ of the DARE \cite[Sec. E.4]{linearestimation}:
\begin{equation}
\begin{aligned}
P
  &= APA^T + Q \\[4pt]
  &\quad - A\tilde{\Gamma}^T H^T
         \bigl(I + H(\tilde{\Gamma} P^{-1} \tilde{\Gamma}^T + \Omega)H^T\bigr)^{-1}
         H\tilde{\Gamma} A^T.
\end{aligned}\label{DARE}
\end{equation}
For $\tilde \Gamma$ to represent a feasible encoder design of a $(p, D)$ code, it is necessary and sufficient that the stabilizing solution $P \succ 0$ of \eqref{DARE} exists and satisfies the power constraint
\begin{equation} \label{base_power}
    \Tr (\tilde \Gamma P^{-1} \tilde \Gamma^T + \Omega) \leq p,
\end{equation}
and MSE constraint
\begin{equation}
    \Tr(P) \leq D.
\end{equation}
\end{lemma}

The addition of the term $M_t$ cannot improve the estimation error since the PSD solution to the DARE \eqref{DARE} increases with respect to the PSD ordering as $\Omega$ increases \cite[Ch. 11]{linearestimation}. Further, $\Omega$ only increases the transmit power in \eqref{base_power}. Thus, any $(p,D)$ achievable with $M_t \neq 0$ is also achievable with $M_t = 0$, and we set $M_t = 0, \forall t$.

\section{Main Results} \label{mainresult}
We present sufficient and necessary conditions for finite MSE achievable with linear encoders in the transmission of a vector-valued source over an arbitrary rank channel. We also investigate whether linear encoders are optimal for MIMO channels.

\subsection{Linear Coding for MIMO Channels}
\subsubsection{Sufficiency}
First, we present a sufficient condition for linear codes to achieve finite MSE.

\begin{theorem}[MIMO channels - sufficiency] \label{thm: MIMO_suff}
\textit{In zero-delay JSCC (Def. \ref{code}) of a $k$-dimensional Gauss-Markov source (Def. \ref{source}) for transmission over an $n$-input MIMO AWGN channel (Def. \ref{channel}) with power constraint $p$, finite asymptotic error, $D < \infty$ \eqref{distortion}, is achievable with linear encoders \eqref{encoding} if there exists an $n$-set partition, $\{\mathcal{S}_i\}_{i=1}^n$, of $\{1, \ldots, k\}$, such that 
\begin{equation} \label{mimo_cond_part}
    \sum_{j \in \mathcal{S}_i} \log |\lambda_j| < C_i(\pi_i), \ \forall i = 1, \ldots, n
\end{equation}
where $\lambda_j$ are the eigenvalues of $A$ (Def. \ref{source}), $C_i(\pi_i) = \frac{1}{2} \log (1 + h_i^2 \pi_i)$ is the Shannon capacity of the $i$th channel with power $\pi_i\ge0$ and 
\begin{equation} \label{powerc}
    \sum_{i=1}^n \pi_i= p.
\end{equation}}
\end{theorem}

The proof is provided in Appendix \ref{app:main}.

\begin{figure}
\begin{center}
    \includegraphics[width=0.45\textwidth]{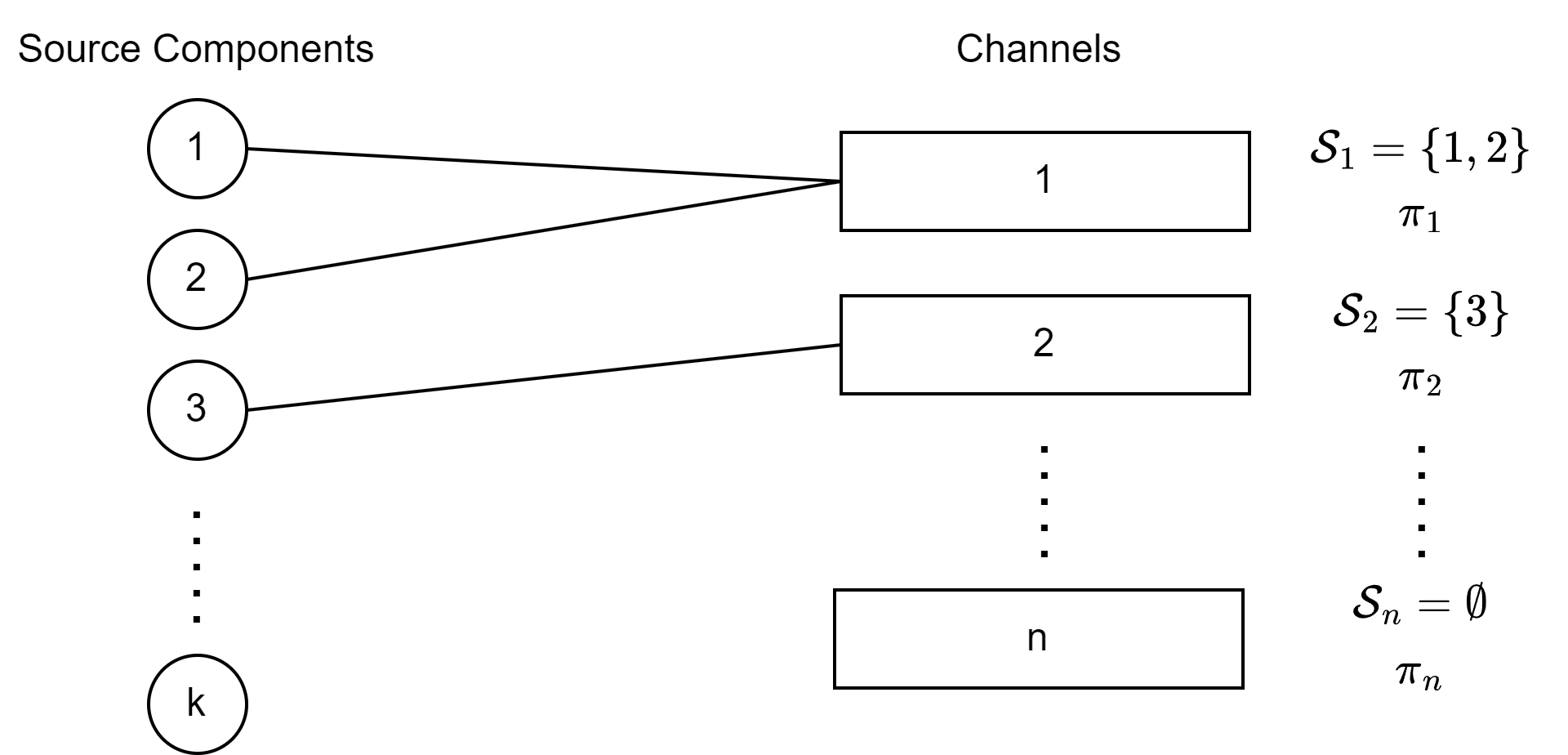}
\end{center}
\caption{Example assignment of source components into the channels and associated power allocations for each channel as described in Theorem \ref{thm: MIMO_suff}. A source can only be allocated to a single channel, and every source must be allocated to a channel. Consequently, the sets $\{\mathcal{S}_i\}_{i=1}^n$ form a partition of $\{1, \ldots, k\}$. The power allocations must satisfy \eqref{powerc} where $p$ is the total power constraint.}
\label{fig: allocation}
\end{figure}

Each set, $\mathcal{S}_i$, of the partition $\{\mathcal{S}_i\}_{i=1}^n$ contains the indices of unstable modes assigned to channel $i$. Thus, Theorem \ref{thm: MIMO_suff} allocates each unstable mode of the source to a single channel output and defines a power allocation over the channels that allows each channel to stabilize its assigned modes independently. See Figure \ref{fig: allocation} for an example.

For an $n$-set partition $\{\mathcal{S}_i\}_{i=1}^n$ and $\pi$ satisfying \eqref{mimo_cond_part} and \eqref{powerc}, we show in Appendix \ref{app:main} that the encoding matrix $\tilde \Gamma = \alpha \Pi^{\frac{1}{2}} \bar \Gamma$, where
\begin{align}
    \Pi &= \text{diag}(\pi_1, \ldots, \pi_n), \\
    \bar \Gamma_{ij} &= \begin{cases}
                        1 & \text{if } j \in \mathcal{S}_i,\\
                        0 & \text{otherwise},
                        \end{cases}
\end{align}
and $\alpha > 0$ is a sufficiently large scaling factor, stabilizes the system with finite MSE and achieves the power constraint \eqref{powerconstraint}. Note that the condition in Theorem \ref{thm: MIMO_suff} does not involve observability conditions that are typical in classical control. This is because we can choose the encoding matrix $\tilde \Gamma$, so that the observation matrix is $\tilde \Gamma H$. Although this condition is difficult to compute as we scale the dimensions of the problem $n, k$, the system designer only needs to perform this computation once for a given system and channel. The application of the resulting estimator to the system is computationally easy, as detailed in Lemma \ref{optimalencoder}.

We show in Theorems \ref{conj: MIMO} and \ref{lem: MIMO_nec}, below, that this partitioning property is necessary if either the source or the channel is scalar, or if Conjecture 1, below, holds.

\subsubsection{Necessity}
It is unknown whether the partitioning property of Theorem~\ref{thm: MIMO_suff} is necessary for vector sources and MIMO channels. We now state a matrix-algebraic conjecture, which, if true, implies the necessity of the partitioning property in Theorem \ref{thm: MIMO_suff}.
\begin{conjecture}[Lyapunov positivity] \label{conj: MIMO_riccati}
    \textit{
    Let $J$ be the unique solution to the Lyapunov equation
    \begin{equation} \label{Jlyap_cond}
        J = B J B - \Gamma^T (I + H \Pi H)^{-1} \Gamma + B \Gamma^T \Gamma B,
    \end{equation}
    where $J \in \mathbb{R}^{k \times k}$ and $\Gamma \in \mathbb{R}^{n \times k}$. $B = \mathrm{diag}(b_1, \ldots, b_k), b_i < 1 \ \forall i$ and $H = \mathrm{diag}(h_1, \ldots, h_n)$ are fixed matrices. For a feasible problem instance (given $B, H$), there exists an optimal solution $\Gamma^*$ to
    \begin{equation}
        \inf_{\substack{\Pi \succeq 0, J \succ 0, \Gamma \colon \\ \eqref{Jlyap_cond}, \ \Pi \text{ is diagonal}}} \Tr(\Pi)
    \end{equation}
    that has exactly one non-zero entry per column.}
\end{conjecture}
We observe Conjecture \ref{conj: MIMO_riccati} holds numerically for all systems we have simulated with $n,k \in \{2,3\}$. An extended discussion of Conjecture \ref{conj: MIMO_riccati} can be found in Appendix \ref{app:main}.
\begin{theorem}[Vector source and MIMO channel - necessity] \label{conj: MIMO}
    \textit{Assume Conjecture \ref{conj: MIMO_riccati} holds. In zero-delay JSCC (\mbox{Def. \ref{code}}) of a vector Gauss-Markov source (Def. \ref{source}) and MIMO AWGN channel (Def. \ref{channel}), finite asymptotic error, $D < \infty$ \eqref{distortion}, is achievable using linear time-invariant encoders \textit{only if} there exists a partition $\{\mathcal{S}_i\}_{i=1}^n$ and power allocation vector $\pi$ satisfying \eqref{mimo_cond_part}, \eqref{powerc} in Theorem \ref{thm: MIMO_suff}.}
\end{theorem}

\subsection{Special Cases}
We detail two cases in which Theorem \ref{conj: MIMO} holds by resolving Conjecture \ref{conj: MIMO_riccati}.
\begin{theorem}[Scalar source or scalar channel] \label{lem: MIMO_nec}
    \textit{If either $n=1$ or $k=1$, finite asymptotic error, $D < \infty$ \eqref{distortion}, is achievable by linear codes if and \textit{only if} there exists a partition $\{\mathcal{S}_i\}_{i=1}^n$ and power allocation vector $\pi$ satisfying \eqref{mimo_cond_part}, \eqref{powerc} in Theorem \ref{thm: MIMO_suff}.}
\end{theorem}

The proofs are provided in Appendix \ref{app: MISO}, \ref{app: SIMO}.

\emph{Case I (Scalar channel):} If the channel is scalar, $H = h_1 \in \mathbb R$, then the conditions of Theorem \ref{thm: MIMO_suff} reduce to $\mathcal{S}_1 = \{1, \ldots, k\}$, $\pi = p$. Then, $D < \infty$ if and only if
\begin{equation} \label{n1cond}
    \sum_{j=1}^k \log |\lambda_j| < \frac{1}{2} \log ( 1 + h_1^2 p).
\end{equation}

In words, a vector-valued Gauss-Markov source can be estimated with finite MSE by linear codes over a scalar AWGN channel if and only if the sum of the logs of unstable eigenvalues of the source is strictly less than the Shannon capacity of the channel.

\emph{Case II (Scalar source):}  If the source is scalar, $A = \lambda_1 \in \mathbb{R}$, then the partition sets $\{\mathcal{S}_i\}_{i=1}^n$ as defined in Theorem \ref{thm: MIMO_suff} satisfy $\mathcal S_i = \{1\}$ and $\mathcal S_l = \emptyset$ for all $l \neq i$. From \eqref{mimo_cond_part}, finite MSE is achievable by linear encoders if and only if
\begin{equation} \label{scalarmimocond}
    \log |\lambda_1| < C_1(p).
\end{equation}
The solution in \eqref{scalarmimocond} allocates all the power to the best available sub-channel. The maximum sub-channel capacity, $C_1(p)$, is smaller than the Shannon capacity $C(p)$ unless the power budget $p$ is low enough so that the classical waterfilling solution which attains the Shannon capacity \cite[Thm. 1]{telatarcapacity} also allocates all the power to the best available sub-channel. Thus, \eqref{scalarmimocond} implies that linear codes are generally suboptimal. See Section \ref{sec:linnotopt} for further discussion on the sub-optimality of linear codes. 

Particularizing Theorem \ref{lem: MIMO_nec} to scalar Gauss-Markov sources that are transmitted over a scalar AWGN channel recovers the classical result from \cite{tatikonda, sahaimitter, Khina} that the finite MSE is achievable if and only if $\log |\lambda_1| < C(p).$ 

\subsection{Proof Outlines of Theorems \ref{thm: MIMO_suff} and \ref{conj: MIMO}} \label{sec: linear}
The complete proofs for Theorems \ref{thm: MIMO_suff} and \ref{conj: MIMO} are presented in Appendix \ref{app:main}. We present the key ideas below.

According to Lemma \ref{lemma: riccati}, which establishes the evolution of the prediction error covariance and its convergent behavior, a finite MSE is achievable if and only if the stabilizing solution $P \succeq 0$ to the DARE \eqref{DARE} exists and satisfies the power constraint \eqref{base_power}. The power constraint \eqref{base_power} can be equivalently understood as the existence of $\Pi \succeq 0$ subject to
\begin{align}
    &\tilde \Gamma P^{-1} \tilde \Gamma^T \preceq \Pi \label{tildeGammapower1},\\
    & \Tr (\Pi) \leq p, \label{power1}
\end{align}
where $P \succ 0$ satisfies the DARE
\begin{align}
    & P = APA^T  + Q - A \tilde \Gamma^T H^T \left(I+H \Pi H^T \right)^{-1}H \tilde \Gamma A^T.\label{Plyap11}
\end{align}
By the Schur complement lemma, \eqref{tildeGammapower1} holds if and only if 
\begin{equation} \label{JMIL}
    J = P - \tilde \Gamma^T \Pi^{-1} \tilde \Gamma \succeq 0.
\end{equation}
By substituting $J$ into \eqref{Plyap11}, we obtain a Lyapunov equation for $J$:
\begin{equation}
\begin{aligned}
& AJA^T - J + A\tilde{\Gamma}^T \Pi^{-1} \tilde{\Gamma} A^T + Q \\[4pt]
&\quad - A\tilde{\Gamma}^T H^T \bigl(I + H\Pi H^T\bigr)^{-1} H\tilde{\Gamma} A^T
       - \tilde{\Gamma}^T \Pi^{-1} \tilde{\Gamma} = 0.
\end{aligned}\label{lyap11}
\end{equation}

We have reduced condition \eqref{tildeGammapower1} that concerns a non-linear DARE \eqref{Plyap11} to constraint \eqref{JMIL} on a Lyapunov equation \eqref{lyap11} that can be analyzed more directly. Our proof techniques are focused on finding conditions for positive semi-definite (PSD) solutions the Lyapunov equation \eqref{lyap11}. In the case when the channel is scalar, $n=1$, the positivity condition simplifies to a condition on the Shannon capacity, which reveals itself on the right side of \eqref{mimo_cond_part} and \eqref{n1cond}.

\section{The Control Setting} \label{sec: control}
\begin{figure*}[!t]
  \centering
  \includegraphics[width=0.9\textwidth]{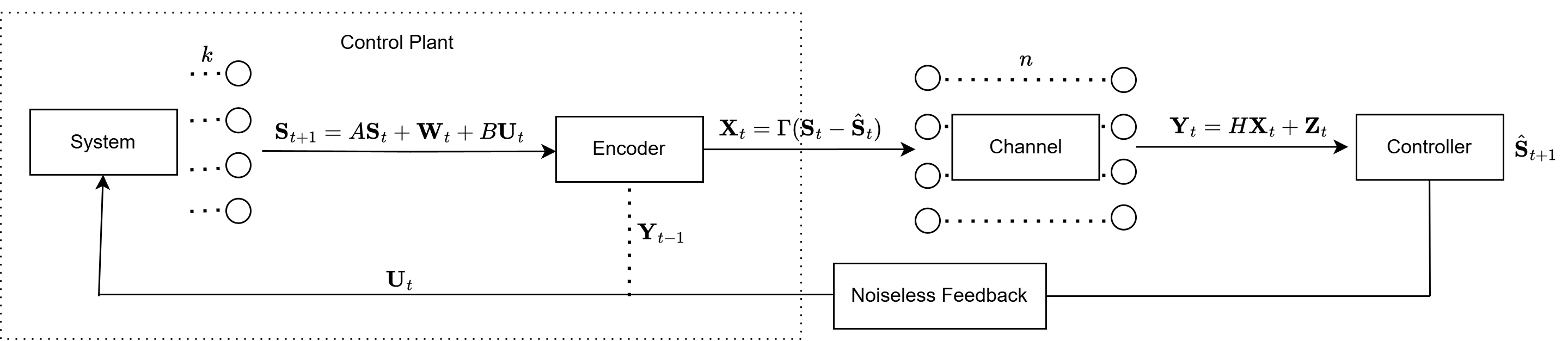}
  \caption{The distributed control setting over a communication channel with
           feedback. The encoder observes the entire state $\mathbf{S}_t$, the
           controller sends the control action noiselessly back to the state,
           and the observed channel output $\mathbf{Y}_{t-1}$ can be inferred
           by the encoder.}
  \label{fig:controlsystem}
\end{figure*}

Our main results translate readily to a remote control problem. The setting is depicted in Figure \ref{fig:controlsystem}. Consider a discrete-time stochastic linear system as in Definition \ref{controlsystem}. The system outputs its recent state $\bfS_t$.
\begin{definition}[Linear dynamical system model]\label{controlsystem}
 The stochastic linear dynamical system evolves according to
\begin{equation} \label{stateupdatecontrol}
    \bfS_{t+1} = A \bfS_t + B \bfU_t + \bfW_t, 
\end{equation}
where $\bfS_t \in \mathbb{R}^k$ is the state, $\bfW_t \in \mathbb{R}^k$ is the process disturbance, $\bfU_t \in \mathbb{R}^l$ is the control signal, and $A \in \mathbb{R}^{k \times k}$,  $B \in \mathbb{R}^{k \times l}$ are fixed matrices. $\bfW_t \sim \mathcal{N}(0,Q)$ is i.i.d and the initial state is $\bfS_0 \sim \mathcal{N} (0, Q)$.
\end{definition}

We use the same technical assumptions---Assumption \ref{asm: controllable}, \ref{asm: diag}, and \ref{asm: time_inv_enc}---as the previous section, and also add Assumption \ref{asm: stabi} which is common in control literature \cite[Apx. C]{linearestimation}.
\begin{assumption} \label{asm: stabi}
    $(A, B)$ is stabilizable and $(A, C^{\frac{1}{2}})$ is detectable.
\end{assumption}

We model the channel as a MIMO AWGN channel as in Definition \ref{channel}. The distinction here is that a controller utilizes the channel outputs, $\bfY^t$ to compute the control action $\bfU_t$. The feedback link carries both the recent channel output $\bfY_t$ and the control action $\bfU_t$ noiselessly back to the system, where the control input is added to the next system state.  

As in the previous section, we assume that the encoder is linear with respect to its inputs. We call a controller \emph{linear} if both the encoder, $\bfX_t = f_t(\bfS^t, \bfY^{t-1}, \bfU^{t-1})$, and controller, $\bfU_{t} = g(\bfY^{t-1})$, are linear functions of their arguments. In the linear-quadratic-Gaussian (LQG) problem, the dimension of the channel matches that of the source, and the system designer does not have the ability to optimize an encoder. In our setting, the encoder must be optimized subject to a power constraint and the channel input dimension is fixed to the input dimension of the MIMO AWGN channel. Furthermore, in the LQG setting, the (noisy) observations are linear combinations of the source states, so if the source blows up, the observation power naturally increases. In our setting, the channel outputs that the controller observes are power constrained to the channel input power constraint. 

The controller seeks to minimize the infinite horizon LQR cost defined as:
\begin{equation} \label{LQR_cost}
    \LQR = \lim_{T\to \infty} \frac{1}{T} \sum_{t=1}^{T} \mathbb{E} \left [  \left( \bfS_t^T C \bfS_t + \bfU_t^T E \bfU_t \right) + \bfS_T^T C \bfS_T\right ],
\end{equation}
where $C \succ 0, E \succ 0$.

We derive conditions for linear encoders to achieve finite LQR cost \eqref{LQR_cost}. This is equivalent to mean-squared stability, meaning
\begin{equation} \label{msstable}
\limsup_{t\to \infty} \E[\bfS_t^T \bfS_t] \leq \infty.
\end{equation}

\subsection{Principle of Separation of Control and Estimation}
It is natural to consider a controller that generates a state estimate $\hat \bfS_t$ and then the control input $\bfU_t$ as if $\hat \bfS_t$ was the true state. We argue such a scheme is optimal. Due to the linear encoder structure, the classical principle of separation of control and estimation holds, resulting in the following Lemma.

\begin{lemma}[Separation principle, {\cite[Thm.~11.2.3]{bb}}] \label{lem: ceq}
Given a source with linear dynamics (Def. \ref{source}) and Gaussian channel (Def. \ref{channel}) with noiseless feedback (``equi-memory"), the LQR control cost can be separated as
\begin{equation} \label{separation}
    \LQR = \Tr(QF) + D,
\end{equation}
where $F$ is the solution to the algebraic Riccati equation
\begin{align}
    F &= C + A^T(F - M)A \\
    M &\triangleq FB(E + B^TFB)^{-1}B^TF.
\end{align}
 In \eqref{separation}, $\Tr(QF)$ is the full-information LQR cost of a controller that observes the system's state $\bfS_t$ directly and $D$ is the infinite-horizon MSE of the source state by the controller \eqref{distortion}. The optimal controller is given by
 \begin{equation}
     U_t = -K\hat\bfS_t,
 \end{equation}
 where $K = (E+B^TFB)^{-1}B^TFA$ and $\bfS_t$ is computed by \eqref{kalmandecode}.
\end{lemma}

\subsection{Translation of Main Results} \label{2-results}

Provided that $\bfU^t$ are available at the encoder at time $t$ via feedback, the classical result of separation holds (Lemma \ref{lem: ceq}). Then, the system in Definition \ref{controlsystem} is mean-squared stabilizable if and only if the controller (decoder) can estimate the source with finite MSE, $D < \infty$ \eqref{distortion}.

\begin{theorem}[Stability] \label{thm:stability}
     In a remote control setting with a $k-$dimensional linear dynamical system (Def. \ref{controlsystem}) over an $n-$input MIMO AWGN Channel (Def. \ref{channel}) with power constraint $p$, the controller attains finite LQR cost and mean-squared stability if and only if $D<\infty$  \eqref{distortion} is achievable for the control-free system and $\Tr(QF) < \infty$ \eqref{separation}.
\end{theorem}
\begin{proof}
    This is a direct consequence of the separation principle: the system is mean-squared stabilizable \eqref{msstable} if and only if $D < \infty$ and $\Tr(QF) < \infty$. The condition $\Tr(QF) < \infty$ holds when $(A, B)$ is stabilizable and $(A, C^{\frac{1}{2}})$ is detectable \cite[Apx. C]{linearestimation}.
\end{proof}

Furthermore, as a consequence of Lemma \ref{lem: ceq}, the controller that achieves the optimal cost, $\LQR$, can separately optimize the full information cost and estimation cost $D$. The LQR cost with full information is minimized by the classical LQR controller, which achieves the value $\Tr(QF)$. By Lemma \ref{optimalencoder}, the optimal linear encoder and decoder that minimize $D$ are given by \eqref{optencoding} and \eqref{kalmandecode}, respectively. Thus, the optimal linear controller uses the given encoder and decoder pair from the estimation problem as the observer followed by the LQR controller.

\section{Source-Channel Matching and Suboptimality of Linear Codes} \label{sec:linnotopt}

\subsection{Relation to the Shannon Capacity and Source-Channel Matching}
The Shannon capacity corresponds to the maximum number of bits (per channel use) that can be reliably distinguished by the receiver in the limit of infinite coding delay. The water-filling solution \cite[Thm. 1]{telatarcapacity} tells us how to best allocate the  transmitter’s power budget for transmitting bits: less transmit power is allocated to noisier sub-channels, while very noisy sub-channels---those whose noise is above the “water level”---are not used for transmitting information at all. The water-filling solution solves the optimization problem
\begin{equation} \label{mimo_shannon_capacity}
    C(p) = \sup_{\substack{\Pi: \Pi \succeq 0 \\ \Tr(\Pi) = p}} \frac{1}{2}  \log \det (I + H \Pi H^T),
\end{equation}
which in the case of a diagonalized channel can be rewritten as 
\begin{equation}
    C(p) = \sup_{\substack{p_i \geq 0 \\ \sum_{i=1}^n p_i = p}} \frac{1}{2} \sum_{i=1}^n \log(1+h_i^2p_i),
\end{equation}
that is, the water-filling solution allocates power to maximize the sum of Shannon capacities of the individual sub-channels.

Our linear encoding scheme in Theorem \ref{thm: MIMO_suff} also allocates power among the subchannels to satisfy the total transmit power constraint \eqref{powerc}. The resulting allocation is generally not waterfilling because \eqref{mimo_cond_part} introduces constraints on the power allocation beyond \eqref{powerc}.


Prior literature, \cite{Nair2} and \cite[Thm. 4.1]{yukselconverse}, give a converse stating that a code, linear or non-linear, achieves finite MSE for a Gauss-Markov source only if
\begin{equation} \label{RCp}
    \sum_{j} \log |\lambda_j| < C(p).
\end{equation}
The condition in our main result \eqref{mimo_cond_part} can be rewritten as
\begin{equation} \label{LSC_opt}
     \sum_{i=1}^n \ \log |\lambda_i| \ < \sup_{\substack{\pi \colon \ \pi_i \geq 0 \\ \eqref{mimo_cond_part}, \ \eqref{powerc}}} \ \sum_{i=1}^n C_i(\pi_i).
\end{equation}
The right hand side of \eqref{LSC_opt} is less than or equal to the Shannon capacity, indicating that the condition for linear encoders is stricter than the general condition in \eqref{RCp}. When they coincide, we say that the source is matched to the channel in the following sense.

\begin{definition}[Source-channel matching for stability]\label{stablematching}
    The source (Def. \ref{source}) and the channel (Def. \ref{channel}) are \emph{matched in the stability sense} at power budget $p \geq 0$ if
    \begin{equation} \label{matching_eq}
         \sum_{i=1}^k \log \ |\lambda_i| \ = \sup_{\substack{\pi \colon \ \pi_i \geq 0 \\ \eqref{mimo_cond_part}, \ \eqref{powerc}}} \ \sum_{i=1}^n C_i(\pi_i) = C(p),
    \end{equation}
    where $C(p)$ is given by \eqref{mimo_shannon_capacity}.
\end{definition}

If a source (Def. \ref{source}) and channel (Def. \ref{channel}) are matched in the stability sense, linear codes require the minimum power to stabilize the system. To see this, consider $p \geq 0$ so that the first equality in \eqref{matching_eq} holds, then by Theorem \ref{thm: MIMO_suff} a linear code can achieve finite MSE for the given source with power budget $p + \epsilon$, for any $\epsilon > 0$. The second equality in \eqref{matching_eq} implies that the sufficiency condition coincides with the converse statement \eqref{RCp}. This converse indicates that no non-linear code can achieve finite MSE with $p' < p$.

The condition in Definition \ref{stablematching} is similar to \cite[Theorem 3.2]{ZAIDI201632} which claims that it is sufficient for linear codes to achieve finite MSE with minimum power budget. Our converse result in Theorem \ref{lem: MIMO_nec} implies that this condition is also necessary.

\subsection{Linear Codes are NOT Optimal for MIMO Channels}
 In this section, we establish, via a counter-example, that linear encoders are generally not optimal for achieving finite MSE over MIMO channels.


\begin{theorem}[Sub-optimality of linear coding] \label{thm: linnotopt}
    There exist source-channel pairs $(A,H,p)$ (Def. \ref{channel}, Def. \ref{source}, \eqref{powerconstraint}) such that all linear codes result in $D = \infty$, while there exist non-linear codes that can achieve $D < \infty$.
\end{theorem}
\begin{proof}
    We show this via a counterexample. Consider the setting of a scalar Gauss-Markov source and a MIMO AWGN channel studied in Theorem \ref{lem: MIMO_nec}. Let $H = I_{2}$ and $p \gg 1$, which correspond to the high SNR regime given two identical channels.
    Shannon-Kotel’nikov mappings as studied in \cite[Thm. 7.1]{Khina} can achieve finite estimation error if 
    \begin{equation}
        \log|\lambda_1| < C(p) - o(1)
    \end{equation}
    as $p \to \infty$, while by Theorem \ref{lem: MIMO_nec}, a linear encoder achieves finite estimation error only if 
    \begin{equation}
        \log|\lambda_1| < \frac{1}{2} \ C(p).
    \end{equation}
    This follows from $\max_i C_i (p) = \frac{1}{2} \log (1+p)$ in \eqref{scalarmimocond} while $C(p) = 2 \cdot \frac{1}{2} \log (1+\frac{p}{2})$ since the water-filling solution places power $\frac{p}{2}$ in each of the parallel channels. Taking the limit $p \to \infty$, $\max_i C_i (p) = \frac{1}{2} \ C(p).$
\end{proof}


\section{Conclusion}
This paper studies sufficient and necessary conditions for linear codes to achieve finite MSE in the transmission of a vector Gauss-Markov source over a MIMO AWGN channel.

In particular, Theorem \ref{thm: MIMO_suff} shows that finite MSE is achievable if one can partition the unstable eigenmodes of the source onto individual sub-channels with appropriate power allocations so that each channel can stabilize its assigned modes independently. Theorem \ref{lem: MIMO_nec} establishes that this allocation is not only sufficient but also necessary when either the source $k=1$ or channel $n=1$ is scalar. Under a matrix-algebraic conjecture (Conj. \ref{conj: MIMO_riccati}), we prove that the same partitioning condition remains necessary in the vector-valued source and MIMO channel setting.

We show (Thm. \ref{thm: linnotopt}) that for some systems and channels, linear codes can be strictly suboptimal: there exist settings (e.g., a scalar source with two identical sub-channels at high SNR) where no linear scheme achieves finite MSE, whereas nonlinear Shannon–Kotelnikov mappings do. In fact, as the SNR grows large, the linear encoder requires twice as much power to achieve finite MSE compared to the non-linear encoder. We propose a notion of source-channel matching in the stability sense in Definition \ref{stablematching} and demonstrate that linear encoders are generally not optimal for MIMO channels in Theorem \ref{thm: linnotopt}.

We demonstrate through the separation of control and estimation (Lem. \ref{lem: ceq}) that all of our above results on finite MSE translate naturally to the control setting to achieve finite LQR cost, or equivalently mean-squared stability of a linear dynamical system (Thm. \ref{thm:stability}).

While we demonstrate the sub-optimality of linear encoders for zero-delay estimation, linear codes remain viable in practice due to their low complexity. An exciting direction for future work is to explore the entire distortion-capacity tradeoff for linear coding of Gauss-Markov sources over AWGN Channels. We believe that the novel analysis of Riccati recursions we develop here can be applied to analyze both finite-horizon LQR costs in the control setting and the rate-cost tradeoffs in the infinite horizon. Although converse bounds for such rate-cost tradeoffs have been analyzed in \cite{kostinahassibirate}, achievability schemes for specific channel models such as the MIMO AWGN channel are poorly understood for such problems.




\bibliographystyle{ieeetr}
\bibliography{ref} 

\appendices

\section{Proof of Theorem \ref{thm: MIMO_suff} and Theorem \ref{conj: MIMO}} \label{app:main}
For both sufficiency and necessity, we start with the DARE \eqref{DARE} and power constraint \eqref{base_power} and show that the feasibility condition on the stabilizing solution of the DARE is equivalent to a condition on a Lyapunov equation. 

\begin{lemma} \label{lem: J_lyap}
    A finite MSE is achievable if and only if there exists a $J \succeq 0, P \succ 0 , \tilde \Gamma, \Pi \succeq 0,  \Tr (\Pi) \leq p$ satisfying
\begin{align}
    \begin{split}
    & AJA^T - J + A \Gamma^T \Pi^{-1} \Gamma A^T + Q \\
    & \qquad \ \ - A \Gamma^T H^T (I + H \Pi H^T)^{-1} H \Gamma A^T\\
    & \qquad \ \ - \Gamma^T \Pi^{-1} \Gamma = 0. \label{J_lyap}\\
    \end{split}
\end{align}
\end{lemma}

\begin{proof}
    First, a finite MSE is achievable if and only if there exists a $P \succ 0, \Gamma, \Omega \succeq 0$ satisfying \eqref{DARE} and \eqref{base_power}. Note that \eqref{DARE} can be translated to a standard DARE with the transformation: $\tilde \Gamma P^{-1} = \Gamma$.
    
    We define $\Pi \triangleq \tilde \Gamma P^{-1} \tilde \Gamma^T + \Omega$ as an auxiliary variable and apply it to \eqref{DARE} and \eqref{base_power}. Now, let $P^+$ denote the solution to \eqref{DARE} when $\Omega \neq 0$ and $P$ denote the solution when $\Omega = 0$. Note that $P^+ \succeq P$, so any $\Omega \neq 0$ hardens the conditions of Lemma \ref{lemma: riccati}. Using this insight, we write the equivalent statement as:

    A finite MSE is achievable if and only if there exists a $P \succ 0 , \tilde \Gamma, \Pi \succeq 0$ satisfying
    \begin{align}
        & P = APA^T  + Q - A \tilde \Gamma^T H^T \left(I+H \Pi H^T \right)^{-1}H \tilde \Gamma A^T \label{Plyap1}\\
        & \Pi \succeq \tilde \Gamma P^{-1} \tilde \Gamma^T \label{tildeGammapower}\\
        & \Tr (\Pi) \leq p.
    \end{align}

Finally, we introduce variables to rewrite the constraints. First,
\begin{equation} \label{Jdef}
    J = P - \tilde \Gamma^T \Pi^{-1} \tilde \Gamma \succeq 0,
\end{equation}
which holds iff \eqref{tildeGammapower} holds by the Schur complement lemma. Then, by substituting $J$ into \eqref{Plyap1}, we obtain a Lyapunov for $J$, given in \eqref{J_lyap}.

Note that $J \succeq 0$ if and only if the constraint \eqref{tildeGammapower} is satisfied by the definition of $J$ in \eqref{Jdef}. The existence of a positive definite solution $P \succ 0$ to the DARE \eqref{Plyap1} is also equivalent to $J \succeq 0$ and \eqref{J_lyap}.
\end{proof}

We will focus our analysis on the feasibility of the constraints in Lemma \ref{lem: J_lyap}. Left and right multiplying \eqref{J_lyap} by $A^{-1}$ and $A^{-T}$ (the inverse exists by Assumption \ref{asm: diag} since the source is strictly unstable) to obtain a stable Lyapunov equation in $J$, we have
\begin{equation}
\begin{aligned}
& A^{-1} J A^{-T} - J - \tilde{\Gamma}^T \Pi^{-1} \tilde{\Gamma} - A^{-1} Q A^{-T} \\[4pt]
&\quad + \tilde{\Gamma}^T H^T \bigl(I + H \Pi H^T\bigr)^{-1} H \tilde{\Gamma} \\[4pt]
&\quad + A^{-1} \tilde{\Gamma}^T \Pi^{-1} \tilde{\Gamma} A^{-T} = 0.
\end{aligned}\label{lyap_J}
\end{equation}
By linearity, we can separate $J = \hat J + \tilde J$ where
\begin{align}
    \hat J &= A^{-1} \hat J A^{-T} - A^{-1}QA^{-T} \\
    \begin{split}
        \tilde J &= A^{-1} \tilde J A^{-T} - \tilde \Gamma^T \Pi^{-1} \tilde \Gamma \\
        & \qquad + \tilde \Gamma^T H^T (I + H \Pi H^T)^{-1}H \tilde \Gamma \\
        & \qquad + A^{-1} \tilde \Gamma^T \Pi^{-1} \tilde \Gamma A^{-T}.
    \end{split}
    \label{tildeJ}
\end{align}
The purpose of this step is to separate the terms involving $\tilde \Gamma$ so that $\tilde \Gamma$ only affects $\tilde J$. If there exists a $\tilde \Gamma$ that makes $\tilde J$ strictly positive, we can arbitrarily scale $\tilde \Gamma$ so that $\tilde J$ is arbitrarily positive. By making $\tilde J$ arbitrarily positive, $J \succeq 0$, so we will limit our investigation to the positivity of $\tilde J$.

Let
\begin{equation}\label{Lambda_def}
    \Pi^{1/2} H^T H \Pi^{1/2} = U\Lambda U^*
\end{equation}
be a singular value decomposition, where $U$ is unitary and $\Lambda$ is diagonal with decreasing positive entries. From \eqref{Lambda_def}, $\Lambda_i = \sigma_i \left( \Pi^{\frac{1}{2}} H^T H \Pi^{\frac{1}{2}} \right)$, where $\sigma_i(\cdot)$ indicates the $i$th singular value in order. We define the notation
\begin{equation} \label{barGamma}
    \overline{\Gamma} \triangleq U\Pi^{-\frac{1}{2}}\tilde \Gamma.
\end{equation}
We now have the following chain of equalities:
\begin{align}
    & \Pi^{-1} - H^T (I + H \Pi H^T)^{-1}H \\
    & = (\Pi + \Pi H^T H \Pi)^{-1} \\ 
    & = \Pi^{-\frac{1}{2}}(I + \Pi^{1/2} H^T H \Pi^{1/2})^{-1} \Pi^{-\frac{1}{2}}\\
    & = \Pi^{-\frac{1}{2}}U^*(I + \Lambda)^{-1} U\Pi^{-\frac{1}{2}}. \label{last_pi}
\end{align}
 Applying \eqref{last_pi} and \eqref{barGamma} to the three latter terms of the right hand side of \eqref{tildeJ}, we have
\begin{align}
    \begin{split}
        & \tilde \Gamma^T \Pi^{-1} \tilde \Gamma - \tilde \Gamma^T H^T (I + H \Pi H^T)^{-1}H \tilde \Gamma  - A^{-1} \tilde \Gamma^T \Pi^{-1} \tilde \Gamma A^{-T}
    \end{split}
    \\
    \begin{split}
        & = \tilde \Gamma^T \Pi^{-\frac{1}{2}}U^*(I + \Lambda)^{-1} U\Pi^{-\frac{1}{2}}\tilde \Gamma  - A^{-1} \tilde \Gamma^T \Pi^{-1} \tilde \Gamma A^{-T} \label{lsvd}\\
    \end{split}
    \\
    \begin{split}
        & = \overline{ \Gamma}^T (I + \Lambda)^{-1} \overline{ \Gamma} - A^{-1} \overline{ \Gamma}^T \overline{ \Gamma} A^{-T}. \label{J_lyap_simp}\\
    \end{split}
 \end{align}
Let $\overline{\Gamma}_i$ denote the $i$th row of $\overline{\Gamma}$. Using \eqref{J_lyap_simp}, we rewrite \eqref{tildeJ} equivalently as
\begin{equation}
     \tilde J = A^{-1} \tilde J A^{-T} + \sum_{i} \left (- \frac1{1 + \Lambda_i}\overline{\Gamma_i}^T \overline{\Gamma_i} + A^{-1} \overline{\Gamma_i}^T \overline{\Gamma_i} A^{-T} \right). \label{Ji_sum_tilde}
 \end{equation}

Using \eqref{Ji_sum_tilde}, we can decompose the Lyapunov equation on $\tilde J$ in the following way: Let $J_i$ be the unique solution to the Lyapunov equation
\begin{equation} \label{LambdaJi}
    \tilde J_i = A^{-1} \tilde J_i A^{-T} - \frac{\bar \Gamma_i^T \bar \Gamma_i}{1 + \Lambda_i} + A^{-1} \bar \Gamma_i^T \bar \Gamma_i A^{-T}.
\end{equation}
Then it follows that the unique solution to $\tilde J$ is given by
\begin{equation} \label{jsum}
    \tilde{J} = \sum J_i.
\end{equation}

To connect the definition of $\bar \Gamma$ to the condition of Theorem \ref{thm: MIMO_suff}, we have the following definition:
\begin{align}
    \begin{split}
        \textit{Let } \mathcal S_i \subset \{1, \ldots, k\} \textit{ be the set of indices where } \\
        \qquad \bar \Gamma_i \textit{ is non-zero.} \label{Si}
    \end{split}
\end{align}
i.e. the source modes indexed by $\mathcal{S}_i$ participate in forming an input to the $i$th sub-channel.

\subsection{Proof of sufficiency: Theorem \ref{thm: MIMO_suff}}
 Let $\tilde J_i^{\mathcal{S}}$ be the sub-matrix of $\tilde J_i$ formed by selecting the column and row indices that are members of $\mathcal{S}_i$, the support of the $i$th row of $\bar \Gamma$. Leveraging the result for vector sources and scalar channels from \eqref{misoji} and \eqref{cap_cond} in Appendix \ref{app: MISO}, below, we conclude that $\tilde J_i^{\mathcal{S}} \succ 0$ if and only if
\begin{equation} \label{MIMO_MISO_cond}
    \sum_{j \in \mathcal{S}_i} \log |a_j| < \frac{1}{2} \log (1 + h_i^2 \pi_i).
\end{equation}
The sufficiency of the theorem follows by letting $\Pi$ be diagonal, in which case $\Lambda_i = \pi_i h_i^2$. Note that $\tilde J_i^{\mathcal{S}}$ is strictly positive provided \eqref{MIMO_MISO_cond} holds, and $\tilde J_i^{\mathcal{S}^c}$, the sub-matrix formed from the rows and columns indexed by the complement, $\mathcal{S}^c$, is zero everywhere. Since the union of supports of the rows of $\bar \Gamma$ include every possible index, $\tilde J = \sum_i \tilde J_i \succ 0$ and can be made arbitrarily positive by scaling $\bar \Gamma$. Such a $\Pi$, $\tilde \Gamma = \Pi^{1/2} \bar \Gamma, J = \tilde J + \hat J, P = J + \tilde \Gamma^T \Pi^{-1} \tilde \Gamma$ are feasible in Lemma \ref{lem: J_lyap}. 

\subsection{Proof of necessity: Theorem \ref{conj: MIMO}}

First, we show that a diagonal $\Pi$ is necessary.

\begin{lemma} \label{lem: diag_pi}
    If $\tilde J$ can be made arbitrarily positive by some $\Gamma, \Pi$, it can also be made arbitrarily positive by $\tilde \Gamma, \tilde \Pi$ where $\tilde \Pi$ is diagonal with $\Tr(\tilde \Pi) \leq \Tr(\Pi)$.
    
    \begin{proof}
    Fix a $\Lambda$ in \eqref{LambdaJi}. We will show that given an arbitrary $\tilde \Gamma, \Pi$ satisfying the power constraint, the same $\Lambda$ is achievable by a diagonal $\tilde \Pi$ and associated $\tilde \Gamma$ with less power.

    By Definition \ref{channel}, $H = \text{diag}\{h_1, \ldots, h_n\}.$
    Without loss of generality, let $h_1 \geq h_2 \geq \ldots \geq h_n$.
    From \eqref{Lambda_def},
    \begin{equation} \label{pi_iden}
        \Pi = U \Lambda^{\frac{1}{2}} V^* (H^T H)^{-1} V \Lambda^{\frac{1}{2}} U^*,
    \end{equation}
    where $V$ is an unitary. To see this,
    take $\Pi^{\frac{1}{2}} = U \Lambda^{\frac{1}{2}}VH^{-1}$ in \eqref{Lambda_def}.
    For a diagonal $\tilde \Pi$,
    \begin{equation} \label{lambda_dec}
        \Lambda = \tilde \Pi^{1/2} H^T H \tilde \Pi^{1/2} = \tilde \Pi H H^T,
    \end{equation}
    since $U = I$ in \eqref{Lambda_def} and $\Pi$, $H$ are both diagonal so they commute. 
     Then by isolating $\tilde \Pi$ and applying the trace to both sides of \eqref{lambda_dec},
    \begin{equation}
        \Tr(\tilde \Pi) = \Tr\left(\Lambda (HH^T)^{-1}\right).
    \end{equation}
    We express the trace of a non-diagonal $\Pi$ in terms of its decomposition in \eqref{pi_iden},
    \begin{align} 
        \Tr(\Pi) &= \Tr(U \Lambda^{\frac{1}{2}} V^* (HH^T)^{-1} V \Lambda^{\frac{1}{2}} U^*) \\
        & = \Tr(\Lambda V^* (HH^T)^{-1} V). \label{notdiagpi}
    \end{align} 
    Note that the entries of $\Lambda$ are in descending order and the entries of $(HH^T)^{-1}$ are in ascending order.

    To compare \eqref{notdiagpi} with the trace of a diagonal $\Pi$, we apply Ruhe's Trace Inequality, which states that if $A,B$ are PSD matrices, with eigenvalues $a_1 \geq a_2 \geq \ldots a_n$ and $b_1 \geq b_2 \geq \ldots b_n$,
    \begin{equation} \label{ruhe}
        \sum_{i=1}^n a_i b_{n-i+1} \leq \Tr(AB).
    \end{equation}

    Applying \eqref{ruhe} with $A = \Lambda$, $B = V^* (HH^T)^{-1} V$, we conclude that $\Tr(\tilde \Pi)$ achieves the lower bound with equality since $\Lambda$ and $(HH^T)^{-1}$ are both diagonal with opposing ordering as in \eqref{ruhe}. Thus,
    \begin{equation}
        \Tr (\tilde \Pi) \leq \Tr (\Pi)
    \end{equation}
    as desired.
\end{proof}    
\end{lemma}

Lemma \ref{lem: diag_pi} shows that we can take $\Pi$ to be diagonal since that is what minimizes the trace of $\Pi$ without affecting the other constraints in Lemma \ref{lem: J_lyap}. Let $h_i$ denote the $i$th diagonal entry of $H$. Then
\begin{equation}
    \Lambda_i = \pi_i h_i^2,
\end{equation}
and
\begin{equation} \label{jisimp}
    \tilde J_i = A^{-1} \tilde J_i A^{-T} - \frac{\bar \Gamma_i^T \bar \Gamma_i}{1 + h_i^2 \pi_i} + A^{-1} \bar \Gamma_i^T \bar \Gamma_i A^{-T}.
\end{equation}

To show the partitioning property of Theorem \ref{thm: MIMO_suff}, we need to show that the sets $\{\mathcal{S}_i\}_{i=1}^n$ cover $\{1, \ldots, k\}$ and are disjoint.

We show the necessity of $\bigcup_{i=1}^k \mathcal{S}_i = \{1, \ldots, k\}$. For $\tilde J$ to be arbitrarily positive, $\bar \Gamma$ must excite all directions of $\tilde J$. To see this, suppose an index $j$ exists such that $\bar \Gamma_{ij} = 0$ for all $i$ in \eqref{barGamma}. Then, $e_j^T \tilde J e_j = 0$, where $e_j$ is the $j$th standard basis vector, and $\tilde J$ cannot be positive definite.

What remains to be shown is the necessity of the disjointness of the sets $\{\mathcal{S}_i\}_{i=1}^n$, which follows from Conjecture \ref{conj: MIMO_riccati}. Indeed, $\tilde J$ \eqref{tildeJ} can be made arbitrarily positive by scaling $\tilde \Gamma$ of the form described in Conjecture \ref{conj: MIMO_riccati}. The sets $\{\mathcal{S}_i\}_{i=1}^n$ are then disjoint.

To provide some intuition regarding Conjecture \ref{conj: MIMO_riccati}, consider the equivalent (redundant) constraints on $P$ in Lemma \ref{lem: MIMO_nec}. Let $\Pi = \tilde \Gamma P^{-1} \tilde \Gamma$, or equivalently, $\Omega = 0$. In Lemma \ref{lem: diag_pi}, we show that the optimal $\Pi$ is diagonal. Consequently, $\tilde \Gamma^T P^{-1} \tilde \Gamma$ is also diagonal. We now ask what this diagonality constraint imposes on the structure of $\tilde \Gamma$.

The original DARE, \eqref{DARE} with $\Omega = 0$ is
\begin{equation}
\begin{aligned}
P
  &= APA^T + Q \\[4pt]
  &\quad - A\tilde{\Gamma}^T H^T
         \bigl(I + H(\tilde{\Gamma} P^{-1} \tilde{\Gamma}^T)H^T\bigr)^{-1}
         H\tilde{\Gamma} A^T.
\end{aligned}\label{st2_DARE}
\end{equation}
When $\tilde \Gamma^T P^{-1} \tilde \Gamma$ is diagonal, the measurement covariance $I + H \tilde \Gamma^T P^{-1} \tilde \Gamma H$ is diagonal as well, since $H$ is diagonal. Conjecture \ref{conj: MIMO_riccati} proclaims that the diagonality of $\tilde \Gamma^T P^{-1} \tilde \Gamma$ imposes a structure on $\tilde \Gamma$ so that $\mathcal{S}_i$ in \eqref{Si} is consistent with the condition of Theorem \ref{thm: MIMO_suff}. This means that $\tilde \Gamma$ has only a single non-zero entry in each column. Equivalently, each mode is assigned to a single channel which implies Theorem \ref{conj: MIMO}.

\section{Proof of Theorem \ref{lem: MIMO_nec} - Vector Source over Scalar Channel} \label{app: MISO}
The channel is scalar, so $n=1$ in \eqref{jisimp}. Considering \eqref{jisimp} with $i=1$ and dropping the subscript for notational simplicity, we have
\begin{equation} \label{misoji}
    \tilde J = A^{-1} \tilde J A^{-T} - \frac{\bar \Gamma^T \bar \Gamma}{1 + h^2 \pi} + A^{-1} \bar \Gamma^T \bar \Gamma A^{-T}.
\end{equation}
Further, we can always set $\pi = p$, which corresponds to using all available power, since increasing $\pi$ increases $\tilde J$ with respect to the PSD ordering.

By Assumption \ref{asm: diag}, $A$ is diagonal. Defining $D_\Gamma \triangleq \mbox{diag}(\bar \Gamma)$, i.e., the diagonal matrix whose components are the elements of the vector $\bar \Gamma^T$, we may now write
\begin{equation} \label{AG_form}
    A^{-1}\bar \Gamma^T = D_\Gamma a ~~~\mbox{and}~~~\bar \Gamma^T = D_\Gamma 1, 
\end{equation}
where $a$ is the vector of the diagonal elements of $A^{-1}$ and $1$ is the all-one vector. Then, (\ref{misoji}) becomes
\begin{equation}
\tilde J = A^{-1}\tilde J A^{-T} + D_\Gamma \left(aa^T-\frac{11^T}{1 + h^2 p} \right)D_\Gamma.
\label{lyap_ru2}
\end{equation}
Let $M$ be the solution to the Lyapunov equation
\begin{equation}
M = A^{-1}MA^{-T}+11^T.
\label{lyap_m}
\end{equation}
The pair $(A, 1)$ is controllable, which holds by Assumption \ref{asm: diag} since the diagonal entries of $A$ are distinct. Then by the Lyapunov stability theorem, $M \succ 0$. We now claim that 
\begin{equation}
{\tilde J} = D_\Gamma\left(\frac{h^2 p}{1 + h^2 p} M -11^T\right)D_\Gamma. 
\label{tilder}
\end{equation}
This can be verified by plugging (\ref{tilder}) into (\ref{lyap_ru2}).
It follows that ${\tilde J} \succ 0$ if and only if $\frac{h^2 p}{1 + h^2 p} M -11^T >0$. But the latter is equivalent to
\begin{equation}
    \left[\begin{array}{cc} M & 1 \\ 1^T & \frac{h^2 p}{1 + h^2 p} \end{array} \right] \succ 0,
\end{equation}
or
\begin{equation}
    \frac{h^2 p}{1 + h^2 p}>1^TM^{-1}1.
\label{power_done}
\end{equation}
Assume $M$ satisfies the Lyapunov equation (\ref{lyap_m}). Then
\begin{equation}
    1^TM^{-1}1 = 1-\left|\mbox{det}(A)\right|^{-2}.
    \label{M_det}
\end{equation}
Indeed, on the one hand
\begin{equation} \label{det_2}
    \mbox{det} \left(A^{-1}MA^{-T}\right) = \left|\mbox{det}A\right|^{-2}\mbox{det}{M},
\end{equation}
and on the other
\begin{equation} \label{det_3}
    \mbox{det}(M-11^T) = (1-1^TM^{-1}1) \ \mbox{det}M,
\end{equation}
and \eqref{M_det} follows since \eqref{det_2} and \eqref{det_3} are equal due to \eqref{lyap_m}.

Then, \eqref{power_done} and \eqref{M_det} imply that $\tilde J \succ 0$ if and only if
\begin{equation}
    \frac{h^2 p}{1 + h^2 p} >1-\left|\mbox{det}A\right|^{-2},
\end{equation}
or equivalently
\begin{equation} \label{cap_cond}
    1 + h^2 p > \left|\mbox{det}A_u\right|^2,
\end{equation}
which is the capacity condition we are seeking.

Note that when this capacity condition holds, $\frac{h^2 p}{1 + h^2 p} M -11^T  \succ 0$ and therefore ${\tilde J} \succ 0$ in \eqref{tilder}. We can arbitrarily scale $\bar \Gamma$ and therefore $D_\Gamma$ to make $\tilde J$ arbitrarily positive. This demonstrates both sufficiency and necessity.

\section{Non-Diagonalizable $A$} \label{app: jordanA}
For a general $A$ that is not diagonalizable, we can extend the proof provided in Appendix \ref{app: MISO}. Consider $A$ in Jordan form: $A = \textrm{diag}(A_1, A_2, \ldots, A_r)$ where $A_i$ represents the Jordan block with eigenvalue $\lambda_i$, and $A \in \mathbb{R}^{k \times k}$. The inverses of the Jordan blocks are given by
\begin{align}
    A_i^{-1} = \lambda_i^{-1}\sum_{k=0}^{m_i-1}(-\lambda_i^{-1})^{\,k}N^{k}, \\
    N \;=\;
  \begin{pmatrix}
    0 & 1 &        &        & 0 \\
      & 0 & \ddots &        &   \\
      &   & \ddots & 1      &   \\
      &   &        & 0      & 1 \\
    0 &   & \cdots & 0      & 0
  \end{pmatrix}
\end{align}
where $m_i$ denotes the multiplicity of eigenvalue $\lambda_i$.

Now, following \eqref{AG_form},
\begin{equation}
    A^{-1} \bar \Gamma^T = D_\Gamma a
\end{equation}
where
\begin{equation}
    D_\Gamma \triangleq \mathrm{diag}(D_1, \ldots, D_r),
\end{equation}
with
\begin{equation}
D_i \;\triangleq\;
\begin{pmatrix}
\Gamma_{i1}      & \Gamma_{i2}      & \Gamma_{i3}      & \cdots & \Gamma_{i\,m_i-1} & \Gamma_{im_i} \\[4pt]
\Gamma_{i2}      & \Gamma_{i3}      & \Gamma_{i4}      & \cdots & \Gamma_{im_i}     & 0           \\[4pt]
\Gamma_{i3}      & \Gamma_{i4}      & \Gamma_{i5}      & \cdots & 0               & 0           \\[2pt]
\vdots           & \vdots           & \vdots           & \ddots & \vdots          & \vdots      \\[2pt]
\Gamma_{im_{i-1}}  & \Gamma_{im_i}      & 0                & \cdots & 0               & 0           \\[4pt]
\Gamma_{im_i}      & 0                & 0                & \cdots & 0               & 0
\end{pmatrix},
\end{equation}
where $\Gamma_{is}$ is the component of $\Gamma$ corresponding to the $s$th coordinate in the $i$th Jordan block, $A_i$, and $s \leq m_i$, $i \leq r$. Also,
\begin{equation}
    a \triangleq (a_1 \ \ldots \ a_r)^T
\end{equation}
where $a_i = (\lambda_i^{-1} \ -\lambda_i^{-2} \ \ldots \ (-1)^{m_i-1} \lambda_i^{-m_i}$).
Further, $\bar \Gamma^T = D_\Gamma e$, where
\begin{equation}
    e \triangleq \bigl(
\underbrace{1,\,0,\dots,0}_{m_{1}\text{ entries}},
\;\underbrace{1,\,0,\dots,0}_{m_{2}\text{ entries}},
\;\dots,
\;\underbrace{1,\,0,\dots,0}_{m_{k}\text{ entries}}
\bigr)^{\!\top}.
\end{equation}
The remainder of the proof follows, with the Lyapunov \eqref{lyap_ru2} becoming

\begin{equation}
\tilde J = A^{-1}\tilde J A^{-T} + D_\Gamma \left(aa^T-\frac{ee^T}{1 + h^2 p} \right)D_\Gamma.
\end{equation}

Similarly, in the proofs of Theorem \ref{thm: MIMO_suff} and Theorem \ref{conj: MIMO} provided in Appendix \ref{app:main}, the $n$-set partition described in \eqref{mimo_cond_part} must not break up the Jordan blocks of $A$. Then, the deconstruction of $\tilde J$ in \eqref{jsum} follows and we can apply the result of vector sources and scalar channels described above without modification in \eqref{MIMO_MISO_cond}.

\section{Proof of Theorem \ref{lem: MIMO_nec} - Scalar Source over MIMO Channel}\label{app: SIMO}
We consider a scalar source and arbitrary rank channel in Theorem \ref{lem: MIMO_nec}, in which case $\tilde \Gamma$ is a column vector and we can solve \eqref{Plyap1} explicitly as 
\begin{equation}
    P = \frac{q - a^2\tilde \Gamma^T H^T (I + H \Pi H^T)^{-1} H \tilde\Gamma}{1-a^2}.
\end{equation}
Plugging this into \eqref{JMIL} we obtain
\begin{equation}
\begin{aligned}
J
  &= -\frac{q}{a^{2}-1} \\
  &\quad
     + \tilde \Gamma^T \left( \frac{a^2}{a^2-1} H^T (I + H \Pi H^T)^{-1} H - \Pi^{-1} \right) \tilde \Gamma \geq 0.
\end{aligned}
\end{equation}

Finite estimation error can be achieved iff there exists a $\tilde \Gamma$ and $\Pi \succeq 0$, $\Tr(\Pi) \leq p$ such that $J \geq 0$. The above statement is equivalent to the statement: finite error is not achievable iff for all $\tilde \Gamma, \Pi$ such that $\Pi \geq 0$, $\Tr (\Pi) \leq p$, we have $J < 0$. This is what we set out to show.

Let
\begin{equation}
    O \triangleq \frac{a^2}{a^2-1} H^T (I + H \Pi H^T)^{-1} H - \Pi^{-1}.
\end{equation}
Note that $J < 0$ for all $\tilde \Gamma$ if and only if $O \preceq 0$, which is equivalent to 
\begin{equation}
    \begin{pmatrix}
        \Pi^{-1} & \sqrt{\frac{a^2}{a^2-1}}H^T \\
        \sqrt{\frac{a^2}{a^2-1}}H & I + H \Pi H^T
    \end{pmatrix} \succeq 0,
\end{equation}
which is also equivalent to
\begin{align}
    I + H\Pi H^T - \frac{a^2}{a^2-1} H \Pi H^T &\succeq 0 \\
    I - \frac{1}{a^2-1} H \Pi H^T & \succeq 0. \label{final_cons}
\end{align}
Recall that $H$ is diagonal, and that $h_1$ is the entry of $H$ with the greatest magnitude. By Lemma \ref{lem: diag_pi}, we can limit $\Pi$ to be diagonal. Thus, \eqref{final_cons} is equivalent to $1-\frac{1}{a^2-1}h_i^2\pi_i \geq 0$ for all $i$, where $\pi_i$ are the diagonal entries of $\Pi$. Consider $\pi_1 = p, \ \pi_2 = \ldots = \pi_n = 0$. Such a $\Pi$ satisfies \eqref{final_cons} if and only if 
\begin{equation} \label{final_res}
    \log |a| \geq \frac{1}{2} \log (1 + h_{1}^2 p).
\end{equation}
If \eqref{final_res} is met, any other $\Pi$ satisfying $\Tr(\Pi) = \sum_i \pi_i \leq p$ will also satisfy $\eqref{final_cons}$.
Thus, \eqref{final_res} is a necessary and sufficient condition for $D< \infty$ to be unattainable. Equivalently, we require 
\begin{equation}
    \log |a| < \frac{1}{2} \log (1 + h_{1}^2 p)
\end{equation}
to achieve $D < \infty$.
\end{document}